\documentstyle[12pt]{article} 
\input psfig.sty

\def\Title#1{\begin{center} {\Large #1 } \end{center}}
\def\Author#1{\begin{center}{ \sc #1} \end{center}}
\def\Address#1{\begin{center}{ \it #1} \end{center}}

\def\doeack{\footnote{Work supported by the Department of Energy,
                     contract DE--AC03--76SF00515.}}
\def\SLAC{Stanford Linear Accelerator Center\\
    Stanford University, Stanford, California 94309 USA}

\newenvironment{Abstract}{\begin{quotation} \begin{center}
                       ABSTRACT
     \end{center}\bigskip  }{\end{quotation}}
\def\beq{\begin{equation}}
\def\eeq#1{\label{#1}\end{equation}}
\def\eeqn{\end{equation}}
\def\beqa{\begin{eqnarray}}
\def\eeqa#1{\label{#1}\end{eqnarray}}
\def\eeqan{\end{eqnarray}}

\def\Re{{\cal R \mskip-4mu \lower.1ex \hbox{\it e}\,}}
\def\Im{{\cal I \mskip-5mu \lower.1ex \hbox{\it m}\,}}
\def\nn{\noindent}
\def\ie{{\it i.e.}}
\def\eg{{\it e.g.}}

\def\etal{{\it et al.}}

\def\sub#1{_{\lower.25ex\hbox{$\scriptstyle#1$}}}
\def\sul#1{_{\kern-.1em#1}}
\def\sll#1{_{\kern-.2em#1}}  
\def\sbl#1{_{\kern-.1em\lower.25ex\hbox{$\scriptstyle#1$}}}
\def\ssb#1{_{\lower.25ex\hbox{$\scriptscriptstyle#1$}}}
\def\sbb#1{_{\lower.4ex\hbox{$\scriptstyle#1$}}}

\def\to{\rightarrow}
\def\mh{\ifmmode m\sbl H \else $m\sbl H$\fi}
\def\mch{\ifmmode m_{H^\pm} \else $m_{H^\pm}$\fi}
\def\mt{\ifmmode m_t\else $m_t$\fi}
\def\mc{\ifmmode m_c\else $m_c$\fi}
\def\mz{\ifmmode M_Z\else $M_Z$\fi}
\def\mw{\ifmmode M_W\else $M_W$\fi}
\def\mws{\ifmmode M_W^2 \else $M_W^2$\fi}
\def\mhs{\ifmmode m_H^2 \else $m_H^2$\fi}   
\def\mzs{\ifmmode M_Z^2 \else $M_Z^2$\fi}
\def\mts{\ifmmode m_t^2 \else $m_t^2$\fi}
\def\mcs{\ifmmode m_c^2 \else $m_c^2$\fi}
\def\mchs{\ifmmode m_{H^\pm}^2 \else $m_{H^\pm}^2$\fi}
\def\ztwo{\ifmmode Z_2\else $Z_2$\fi}
\def\zone{\ifmmode Z_1\else $Z_1$\fi}
\def\mtwo{\ifmmode M_2\else $M_2$\fi}
\def\mone{\ifmmode M_1\else $M_1$\fi}
\def\tb{\ifmmode \tan\beta \else $\tan\beta$\fi}
\def\xw{\ifmmode x\sub w\else $x\sub w$\fi}
\def\ch{\ifmmode H^\pm \else $H^\pm$\fi}
\def\lum{\ifmmode {\cal L}\else ${\cal L}$\fi}
\def\inpb{\ifmmode {\rm pb}^{-1}\else ${\rm pb}^{-1}$\fi}
\def\infb{\ifmmode {\rm fb}^{-1}\else ${\rm fb}^{-1}$\fi}
\def\epem{\ifmmode e^+e^-\else $e^+e^-$\fi}
\def\ppb{\ifmmode \bar pp\else $\bar pp$\fi}

\def\bsg{\ifmmode b\rightarrow s\gamma \else $b\rightarrow s\gamma$\fi}
 
\newskip\zatskip \zatskip=0pt plus0pt minus0pt
\def\matth{\mathsurround=0pt}

\def\atversim#1#2{\lower0.7ex\vbox{\baselineskip\zatskip\lineskip\zatskip
  \lineskiplimit 0pt\ialign{$\matth#1\hfil##\hfil$\crcr#2\crcr\sim\crcr}}}

\begin{document}
\rightline{\vbox{\halign{&#\hfil\cr
&SLAC-PUB-7297\cr
&September 1996\cr}}}
\vspace{0.8in} 
\Title{Constraints on $q\bar q \gamma\gamma$ Contact Interactions at Future 
Hadron Colliders}
\bigskip
\Author{Thomas G. Rizzo\doeack}
\Address{\SLAC}
\bigskip
\begin{Abstract}
We explore the capability of the Tevatron and LHC as well as other future 
hadron colliders to place limits on the possible existence of 
flavor-independent $q \bar q \gamma\gamma$ contact interactions which can 
lead to an excess of high $p_t$ diphoton events with large invariant masses. 
Constraints on the corresponding $e^+e^-\gamma\gamma$ contact interaction 
already exist from LEP. In the case of hadron colliders, strong constraints 
on the scale associated with such interactions are achievable in all cases, 
\eg, of order 0.9(3) TeV at TeV33(LHC). 
\end{Abstract}
\bigskip
\vskip1.0in
\begin{center}
To appear in the {\it Proceedings of the 1996 DPF/DPB Summer Study on New
 Directions for High Energy Physics-Snowmass96}, Snowmass, CO, 
25 June-12 July, 1996. 
\end{center}
%
\bigskip
\def\thefootnote{\fnsymbol{footnote}}
\setcounter{footnote}{0}
\newpage

\section{Introduction}

Although the Standard Model(SM) appears to be as healthy as 
ever{\cite {blondel}}, 
it is generally believed that new physics must exist to address all of the 
questions the SM leaves unanswered and which can explain the values of the 
various input parameters (\eg, fermion masses and mixing angles). Although 
there are many suggestions in the literature, no one truly knows the form 
this new physics might take or how it may first manifest itself. Instead of 
the direct production of new particles, physics beyond the SM may first 
appear as deviations in observables away from SM expectations, such as in the 
rates for rare processes or in precision electroweak tests. Another 
possibility is that deviations of order unity may be observed in cross 
sections once sufficiently high energy scales are probed. This kind of new 
physics can generally be parameterized via a finite set of non-renormalizable 
contact interactions, an approach which has been quite popular in the 
literature{\cite {contact}} for many years. 

In this paper we will explore the capability of both the Tevatron and LHC, as 
well as possible future $\sqrt s$=60 and 200 TeV hadron colliders, to probe 
for the existence of flavor-independent(apart from electric charge),   
$q \bar q \gamma\gamma$ contact interactions of dimension-8. This is the 
lowest dimension gauge invariant operator involving two fermions as well as 
two photons. The observation of the signatures associated with the existence 
of this operator, which are discussed below, would be a clear signal of 
compositeness. Searches for such operators, with the 
quarks replaced by electrons, \ie, $e^+e^-\gamma\gamma$ contact interactions, 
have already been performed at LEP1.5 
{\cite {lep}} and have resulted in a lower bound of approximately 170 GeV 
on the associated mass scale. This constraint is not far above LEP's center 
of mass energy. Since this possibility has been  
explored for the Tevatron and LHC previously{\cite {tgr}}, we 
refer the reader to earlier work for details in these two cases. 
As we will see below, the Tevatron(LHC) with an integrated luminosity of 
2(100) $fb^{-1}$ will easily be able to push the scale in the corresponding 
$q \bar q \gamma\gamma$ situation 
above to $\simeq 0.7(2.8)$ TeV with even larger scales accessible at the higher 
energy hadron machines or with increased luminosity. 

To be definitive, we will follow the notation employed by {\cite {con}} as 
well as by the LEP collaborations and assume that these 
new interactions are parity and $CP$ conserving. In this case we can 
parameterize the lowest dimensional $q \bar q \gamma\gamma$ contact 
interaction as 
\begin{equation}
{\cal L}= {2ie^2\over {\Lambda^4}}Q_q^2 F^{\mu\sigma}F_\sigma^\nu \bar q 
\gamma_\mu\partial_\nu q  \,,
\end{equation}
where $e$ is the usual electromagnetic coupling, $Q_q$ is the quark charge, and 
$\Lambda$ is the associated mass scale. Note that we have pulled out an 
overall factor of $e^2$ as this represents the strength of the new 
interaction associated with the couplings of two photons to a pair of 
fermions, however one is not forced to follow this convention. We have chosen 
this particular form of the interaction as to be able 
to directly compare with the limits obtained at $e^+e^-$ colliders. 
The most obvious manifestation of this 
new operator is to modify the conventional Born-level partonic $q \bar q \to 
\gamma\gamma$ differential cross section so that it now takes the form 
\begin{eqnarray}
{d\hat \sigma\over {dz}}& = & Q_q^4 {2\pi \alpha^2\over {3\hat s}}
\left[{1+z^2\over {1-z^2}} \pm 2{\hat s^2\over {4\Lambda_{\pm}^4}}(1+z^2)
\right. \nonumber\\
& & \left. +\left({\hat s^2\over {4\Lambda_{\pm}^4}}\right)^2 
(1-z^4)\right]  \,,
\end{eqnarray}
where $\hat s,~z$ are the partonic center of mass energy and the cosine of 
center of mass scattering angle, $\theta^*$, respectively. Note that we 
have written $\Lambda_{\pm}$ in place of $\Lambda$ in the equation above 
to indicate that the limits we obtain below will depend upon whether 
the new operator constructively or destructively interferes with the SM 
contribution. This is also reflected by the choice of sign in the cross term 
in the above expression for the parton-level cross section. 

\vspace*{-0.5cm}
\nn
\begin{figure}[htbp]
\centerline{
\psfig{figure=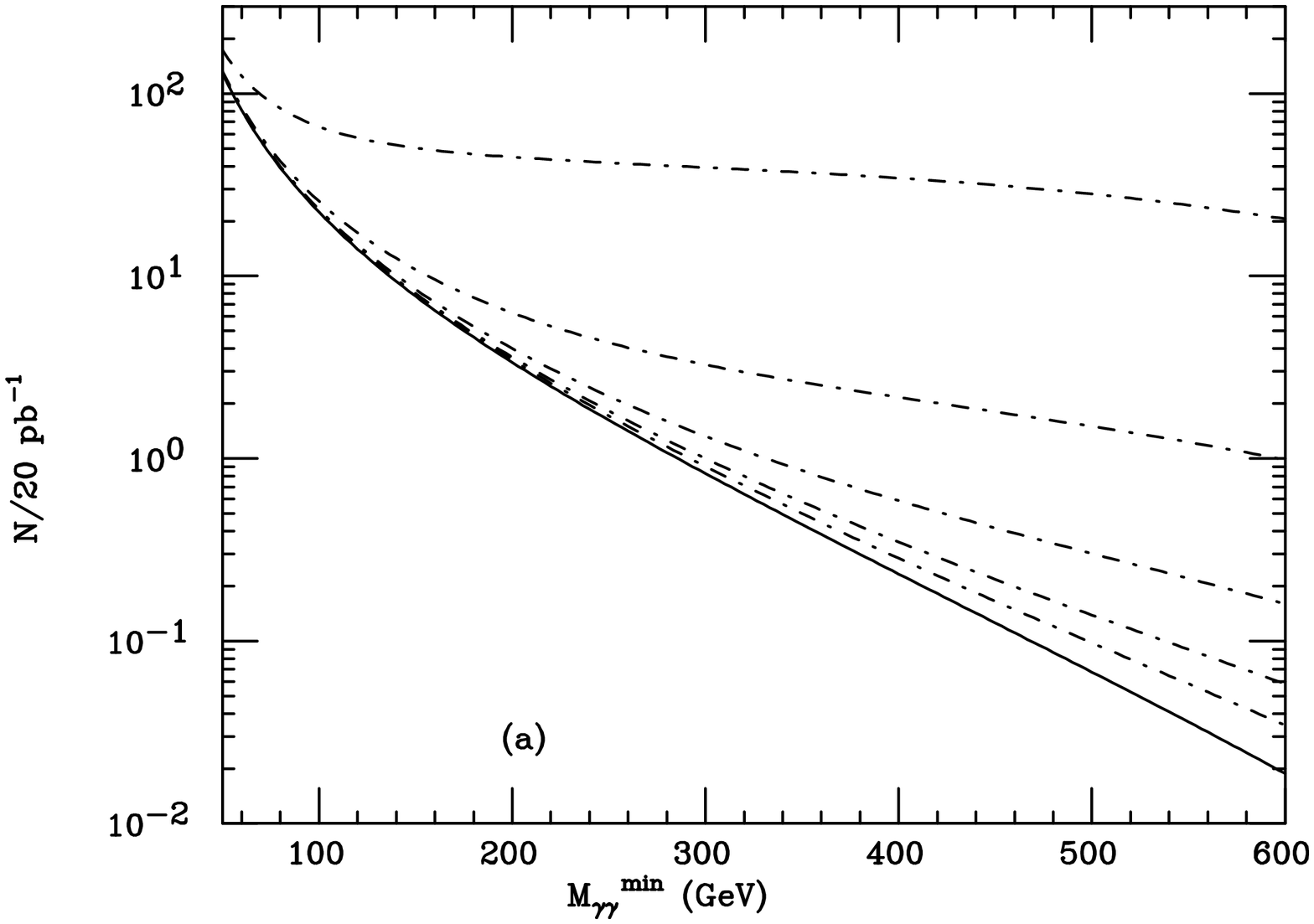,height=9.1cm,width=9.1cm,angle=90}
\hspace*{-5mm}
\psfig{figure=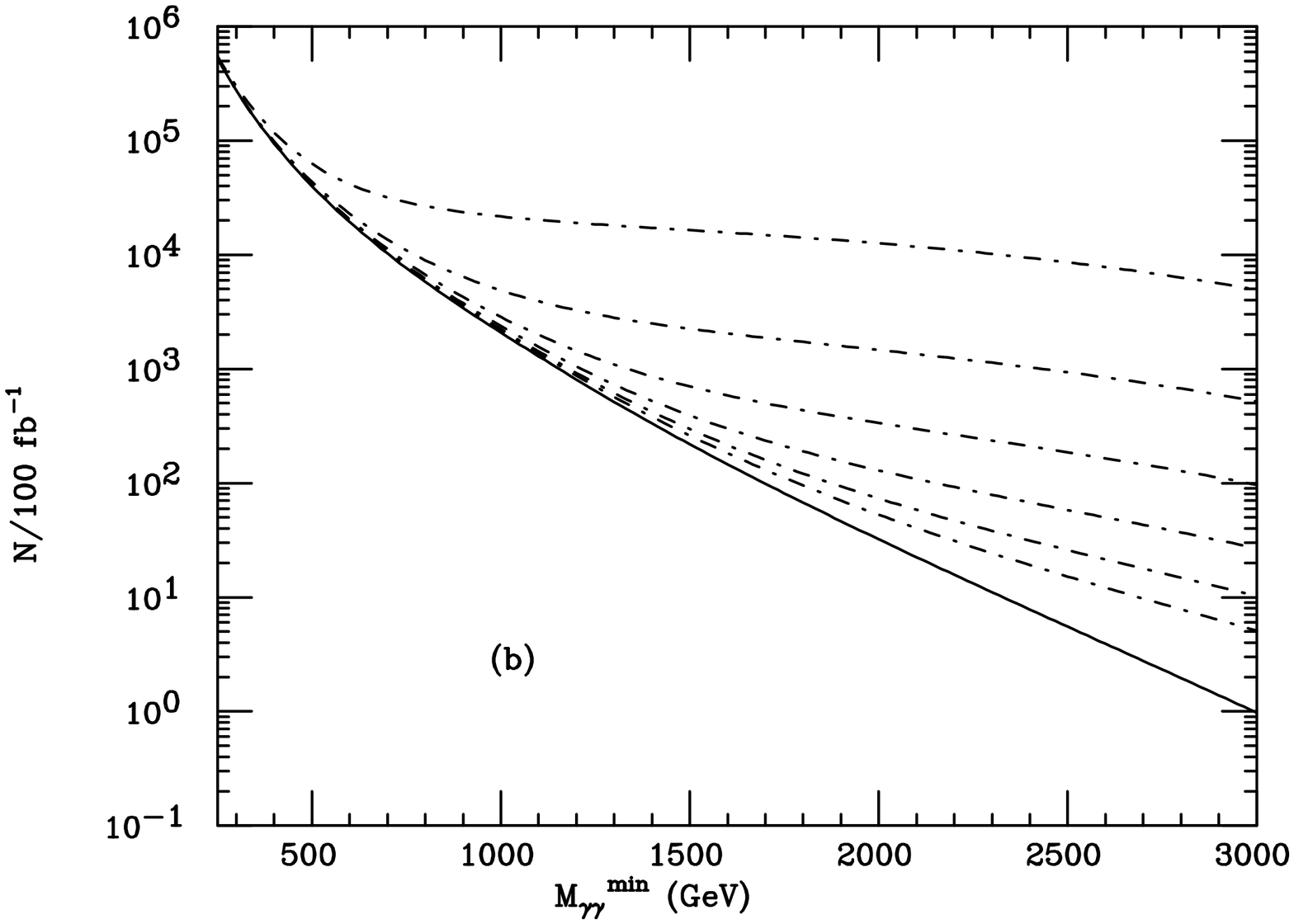,height=9.1cm,width=9.1cm,angle=90}}
\vspace*{-0.6cm}
\caption{(a)Diphoton pair event rate, scaled to an integrated 
luminosity of $20 pb^{-1}$, as a function of $M_{\gamma\gamma}^{min}$ at the 
1.8 TeV Tevatron subject to the cuts $p_t^{\gamma}>15$ GeV 
and $|\eta_\gamma|<1$. The solid curve is the 
QCD prediction, while from top to bottom the dash dotted curves correspond to 
constructive interference with the SM and 
a compositeness scale associated with the $q\bar q \gamma\gamma$ operator of 
$\Lambda_+=0.2,~0.3,~0.4,~0.5,$ and 0.6 TeV respectively. 
(b) Same as (a), but for the LHC scaling to an integrated 
lumonosity of $100 fb^{-1}$. From top to bottom the 
dash dotted curves now correspond to $\Lambda_+=0.75,~1.0,~1.25,~1.5,~1.75$ and 
2.0 TeV respectively. Here we require instead $p_t^{\gamma}>200$ GeV 
and $|\eta_\gamma|<1$.}
\label{tevlhc}
\end{figure}
\vspace*{0.1mm}

\section{Contact Interaction Effects}

There are two major 
effects due to finite $\Lambda$: ($i$) Clearly, once $\hat s$ 
becomes even remotely comparable to $\Lambda^2$, the parton-level diphoton 
differential cross section becomes less 
peaked in the forward and backward directions implying that the photon pair 
will generally be more central and will occur with higher average values 
of $p_t$. 
($ii$) When integrated over 
parton distributions the resulting cross section will lead to an increased 
rate for photon pairs with large $\gamma\gamma$ invariant masses. From 
these two observations we 
see that the best hope for isolating finite $\Lambda$ contributions is to look 
for excess diphotons with high, balanced $p_t$'s in the central detector 
region with large pair masses. To take advantage of the fact that the 
sensitivity to the new contact interaction is greatest in the cental region 
we will demand both photons satisfy $|\eta_\gamma| \leq 1$ and apply 
increasingly strong $p_t$ cuts on both photons as the collider center of mass 
energy increases. We will then examine the sensitivity to finite $\Lambda_\pm$ 
as a function of a lower cut placed on the $\gamma\gamma$ invariant 
mass, $M_{\gamma\gamma}^{min}$. This follows the basic procedure in our 
earlier analysis in Ref.{\cite {tgr}}.

\vspace*{-0.5cm}
\nn
\begin{figure}[htbp]
\centerline{
\psfig{figure=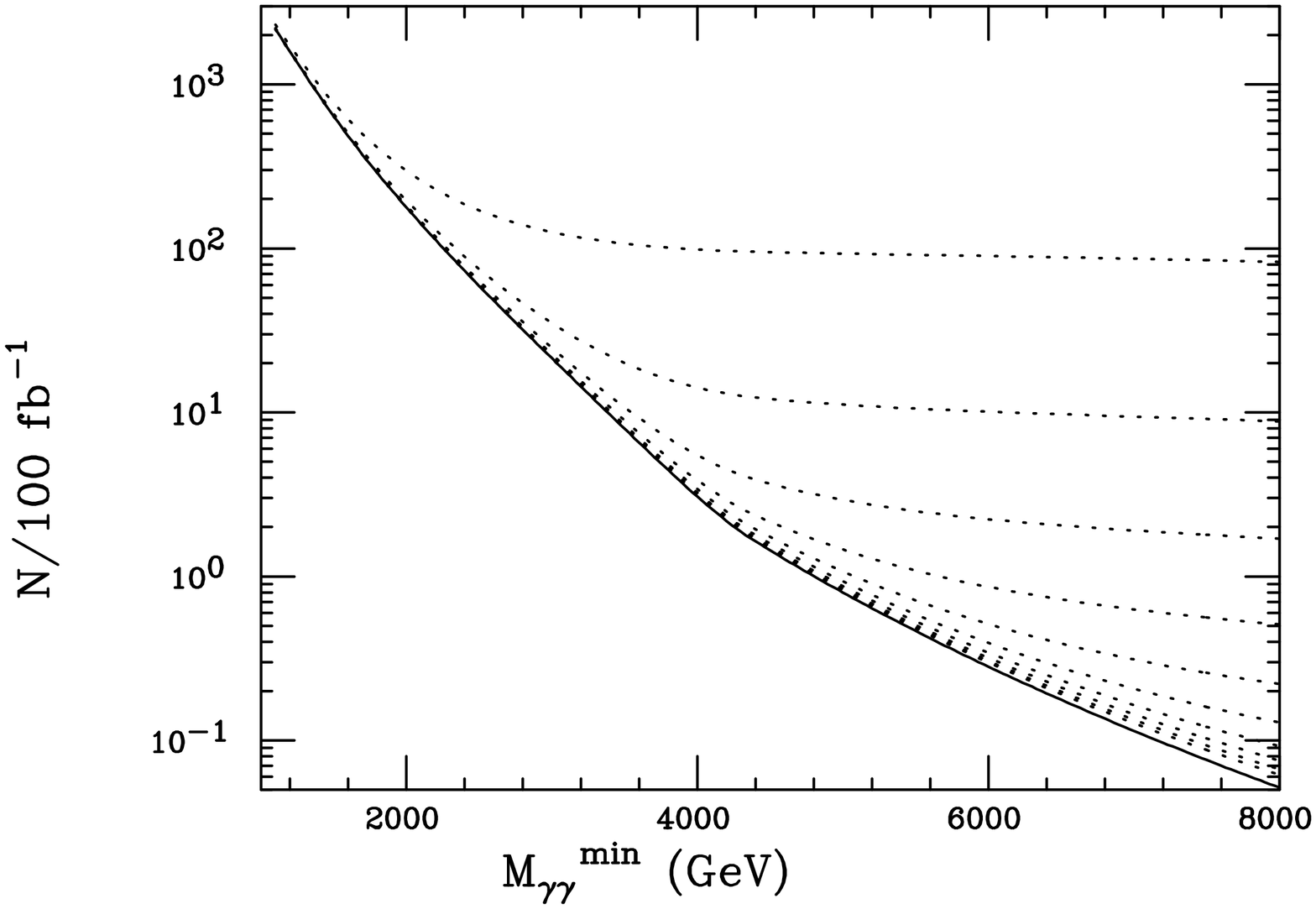,height=9.1cm,width=9.1cm,angle=-90}
\hspace*{-5mm}
\psfig{figure=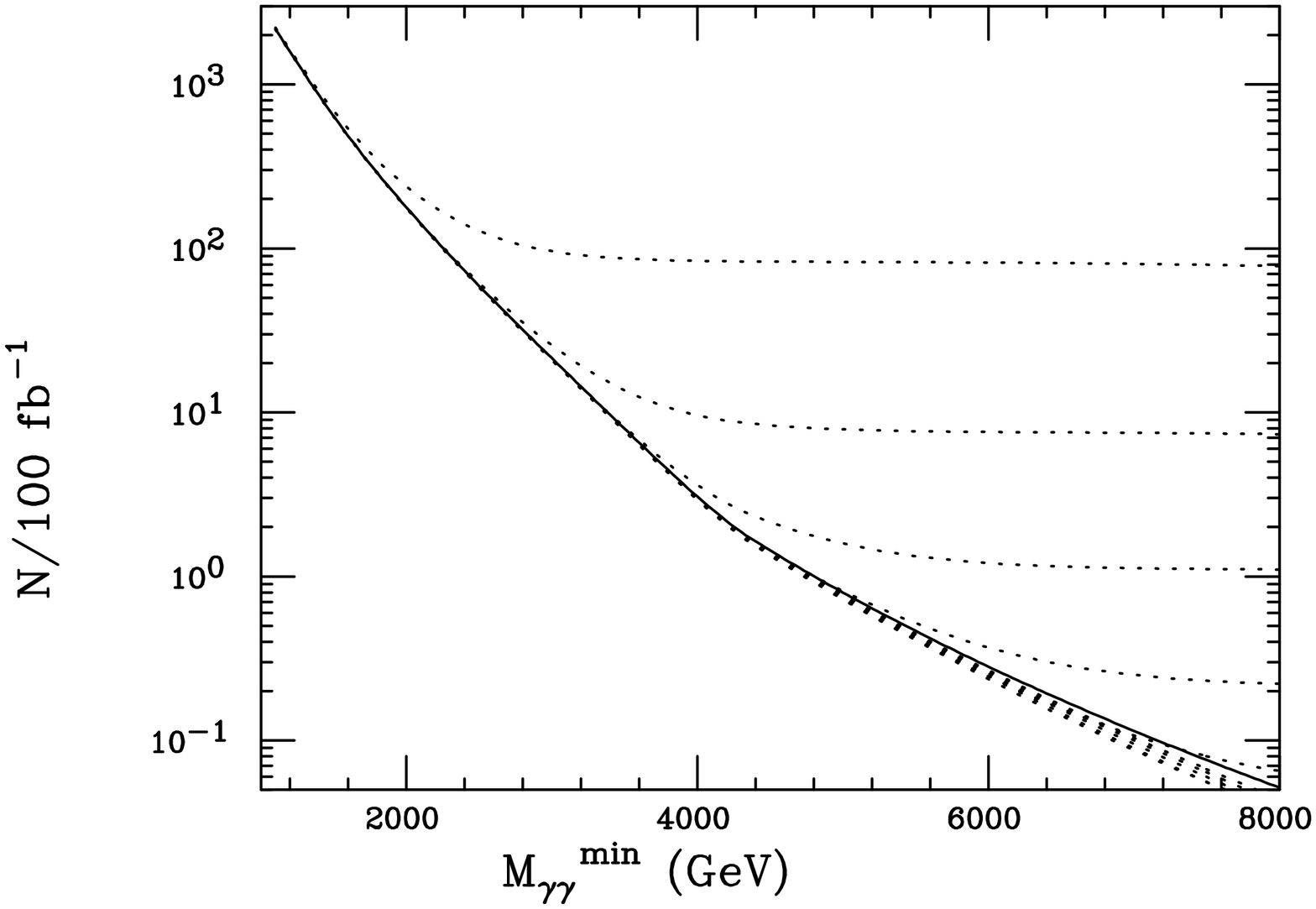,height=9.1cm,width=9.1cm,angle=-90}}
\vspace*{-0.6cm}
\caption{Event rate for isolated $\gamma \gamma$ events with invariant masses 
larger than $M_{\gamma\gamma}^{min}$ at a 60 TeV $pp$ 
collider scaled to a luminosity of 100 $fb^{-1}$. 
The solid curve is the SM case while the top dotted curve corresponds to 
$\Lambda_+(\Lambda_-)=3$ TeV in the left(right) figure. Each subsequent dotted 
curve corresponds to an increase in $\Lambda_{\pm}$ by 1 TeV. In either case 
we have applied the cuts $p_t^\gamma \geq 500$ GeV and $|\eta_\gamma| \leq 1$.}
\label{60tev1}
\end{figure}
\vspace*{0.1mm}

Unfortunately, the $q \bar q \to \gamma\gamma$ tree-level 
process is not the only one which produces diphotons that can satisfy the above 
criteria. The authors of 
Ref.{\cite {nlo}} have provided an excellent summary of the various sources 
which lead to diphoton pairs and we will generally follow their discussion. 
The most obvious additional source of diphotons arises from the process 
$gg\to \gamma\gamma$ which is induced by box diagrams. Although relatively 
small in rate at the Tevatron, the increased $gg$ luminosity as one goes 
to LHC (or higher) energies, combined with the fact that $q\bar q$ annihilation 
is now a `sea-times-valence' process at $pp$ colliders, implies that the 
subprocess $gg\to \gamma\gamma$ will be extremely important there. 

We include the $gg-$induced diphotons in our calculations by 
employing 5 active quark flavors in the $gg$-induced box diagram for partonic 
center of mass energies, $\hat s$, below $4m_t^2$, 6 active flavors for 
$\hat s >>4m_t^2$, and smoothly interpolate between these two cases. In 
order to include the potential 
effects of loop corrections to the rate for $gg\to \gamma\gamma$, we have 
scaled the results obtained by this procedure by an approximate `K-factor' of 
1.3(1.5) at the Tevatron(LHC and higher energy colliders). 
A similar `K-factor' is also employed in the 
$q\bar q \to \gamma\gamma$ calculation; we use the results of Barger, Lopez 
and Putikka in Ref.{\cite {vb}}. While this procedure gives only an 
approximate result in comparison to 
the full NLO calculation, it is sufficient for our purposes since the effects 
of the new contact interaction are quite large as they directly modify the 
tree level cross section. 

\vspace*{-0.5cm}
\nn
\begin{figure}[htbp]
\centerline{
\psfig{figure=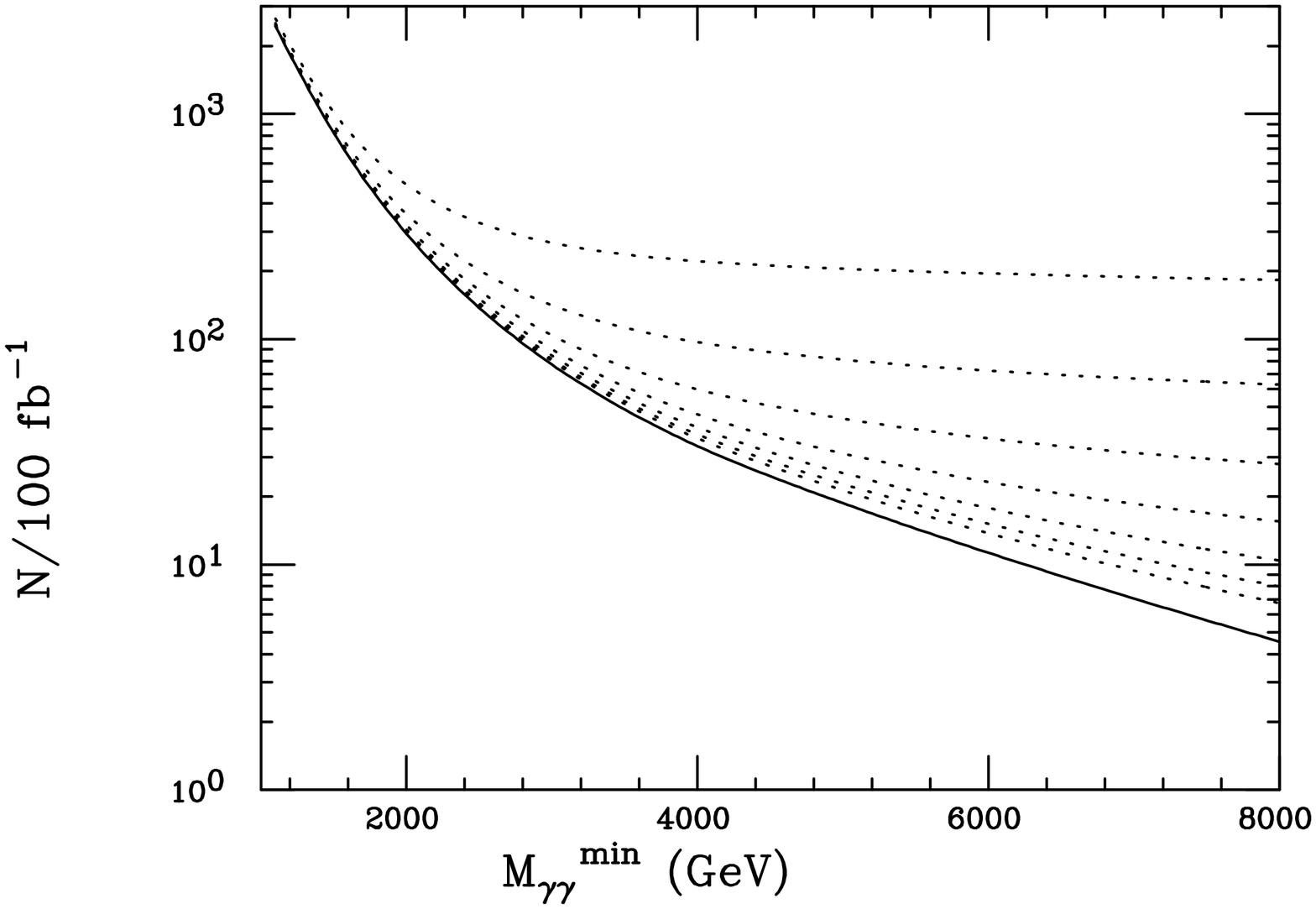,height=9.1cm,width=9.1cm,angle=-90}
\hspace*{-5mm}
\psfig{figure=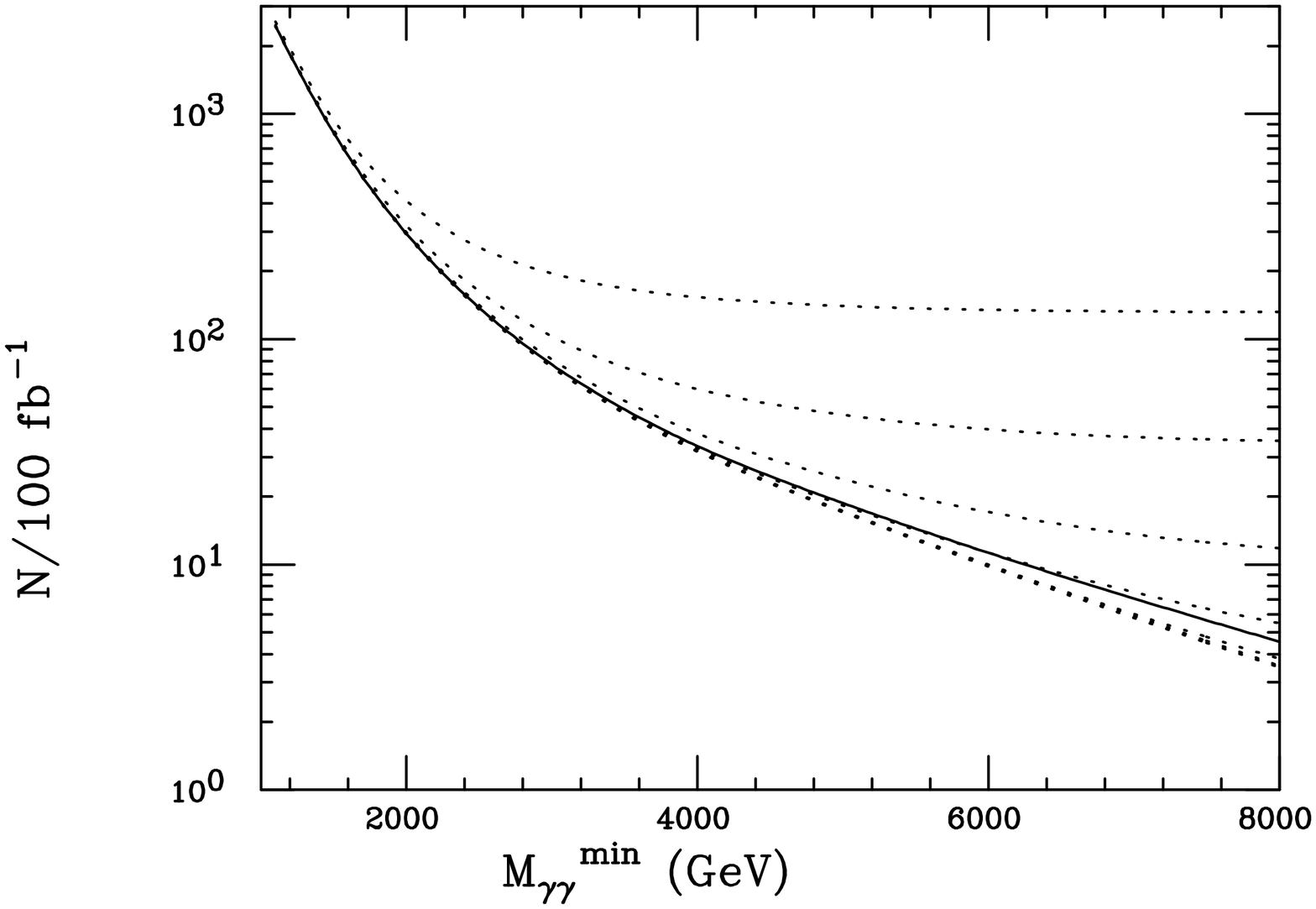,height=9.1cm,width=9.1cm,angle=-90}}
\vspace*{-0.6cm}
\caption{Same as the previous figure but now for the $p\bar p$ collision mode. 
The solid curve is the SM case while the top dotted curve corresponds to 
$\Lambda_+(\Lambda_-)=8$ TeV in the left(right) figure. Each subsequent dotted 
curve corresponds to an increase in $\Lambda_{\pm}$ by 1 TeV.}
\label{60tev2}
\end{figure}
\vspace*{0.1mm}

Additional `background' diphotons arise from three other sources. 
For $2\to 2$ processes, 
one can have either ($i$) single photon production through $gq \to \gamma q$ 
and/or  $q\bar q\to g\gamma$ followed by the fragmentation $g,q \to \gamma$, or 
($ii$) a conventional $2\to 2$ process with {\it both} final state $q,g$ 
partons fragmenting to photons. Although these processes appear to be 
suppressed by powers of $\alpha_s$, these are off-set by large logs. 
For $2\to 3$ processes, ($iii$) double bremsstrahlung production of diphotons 
is possible, \eg, $gq \to q\gamma \gamma$ or $q\bar q \to g\gamma \gamma$. All 
of these `backgrounds' are relatively easy to drastically reduce or completely 
eliminate by a series of isolation cuts and by demanding $p_t$ balancing 
between the two photons, which we require to be back to back in their center 
of mass frame. 

\vspace*{-0.5cm}
\nn
\begin{figure}[htbp]
\centerline{
\psfig{figure=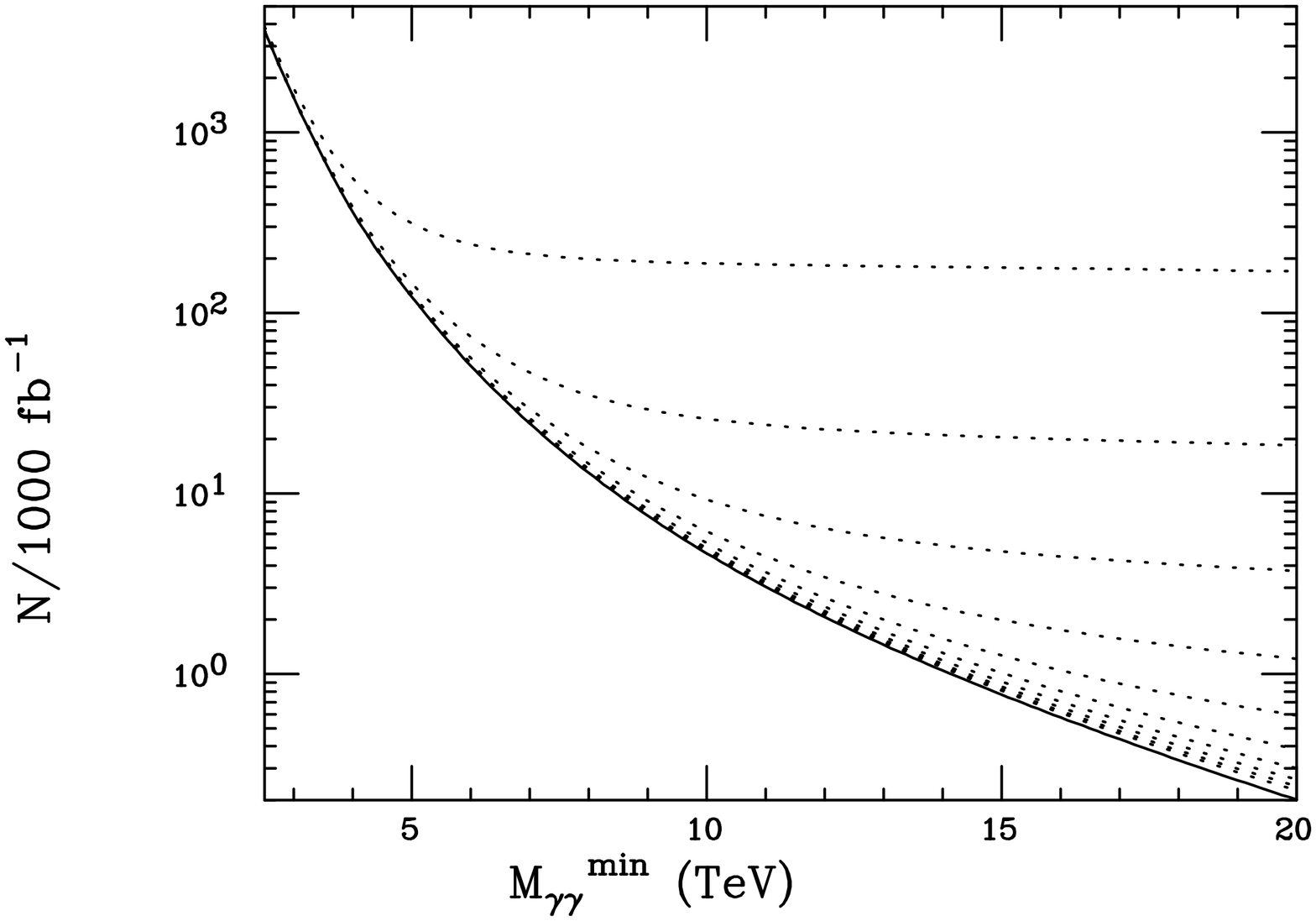,height=9.1cm,width=9.1cm,angle=-90}
\hspace*{-5mm}
\psfig{figure=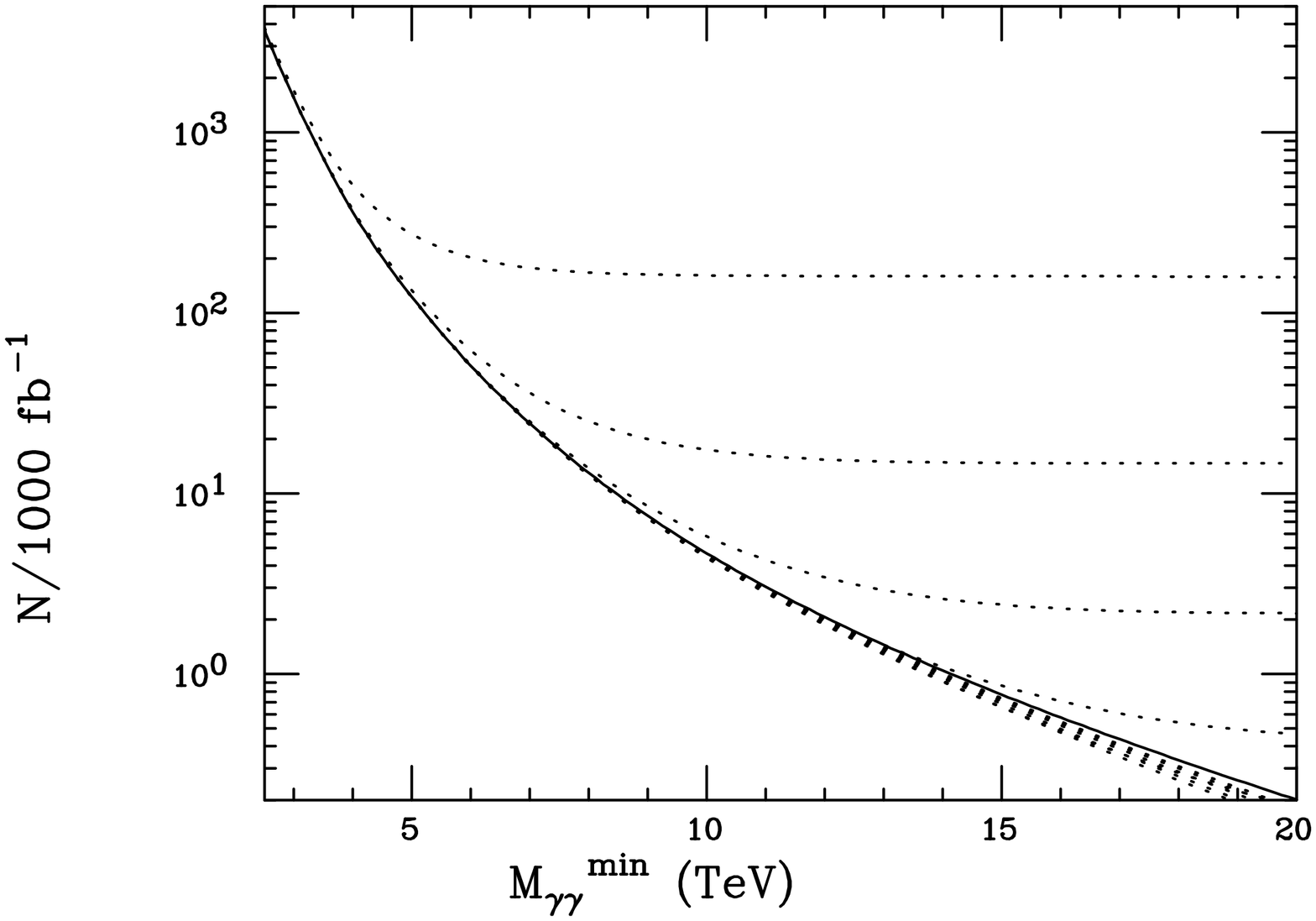,height=9.1cm,width=9.1cm,angle=-90}}
\vspace*{-0.6cm}
\caption{Event rate for isolated $\gamma \gamma$ events with invariant masses 
larger than $M_{\gamma\gamma}^{min}$ at a 200 TeV $pp$ 
collider scaled to a luminosity of 1000 $fb^{-1}$. 
The solid curve is the SM case while the top dotted curve corresponds to 
$\Lambda_+(\Lambda_-)=6$ TeV in the left(right) figure. Each subsequent dotted 
curve corresponds to an increase in $\Lambda_{\pm}$ by 3 TeV. In either case 
we have applied the cuts $p_t^\gamma \geq 1$ TeV and $|\eta_\gamma| \leq 1$.}
\label{200tev1}
\end{figure}
\vspace*{0.1mm}

\section{Results}

To obtain our results we fold the two parton-level subprocess cross sections 
with their 
associated parton densities, scale by appropriate K-factors and machine 
luminosities and then integrate 
over the relevant kinematic domain, subject to the appropriate cuts on both 
$p_t$ and $\eta_{\gamma}$. We then obtain the number of events with the 
diphoton pair invariant mass larger than $M_{\gamma\gamma}^{min}$, which we 
display for different values of $\Lambda_{\pm}$ appropriate to the collider's  
center of mass energy. 
The results for the Tevatron and LHC can be seen in Fig.~\ref{tevlhc}. 
Fig.~\ref{tevlhc}a compares the SM diphoton cross section as a function of
$M_{\gamma\gamma}^{min}$ with the constructive interference scenario for 
various values of $\Lambda$. It is clear that present data 
from the Tevatron is already probing values of $\Lambda$ of order 400-500 GeV.

\vspace*{-0.5cm}
\nn
\begin{figure}[htbp]
\centerline{
\psfig{figure=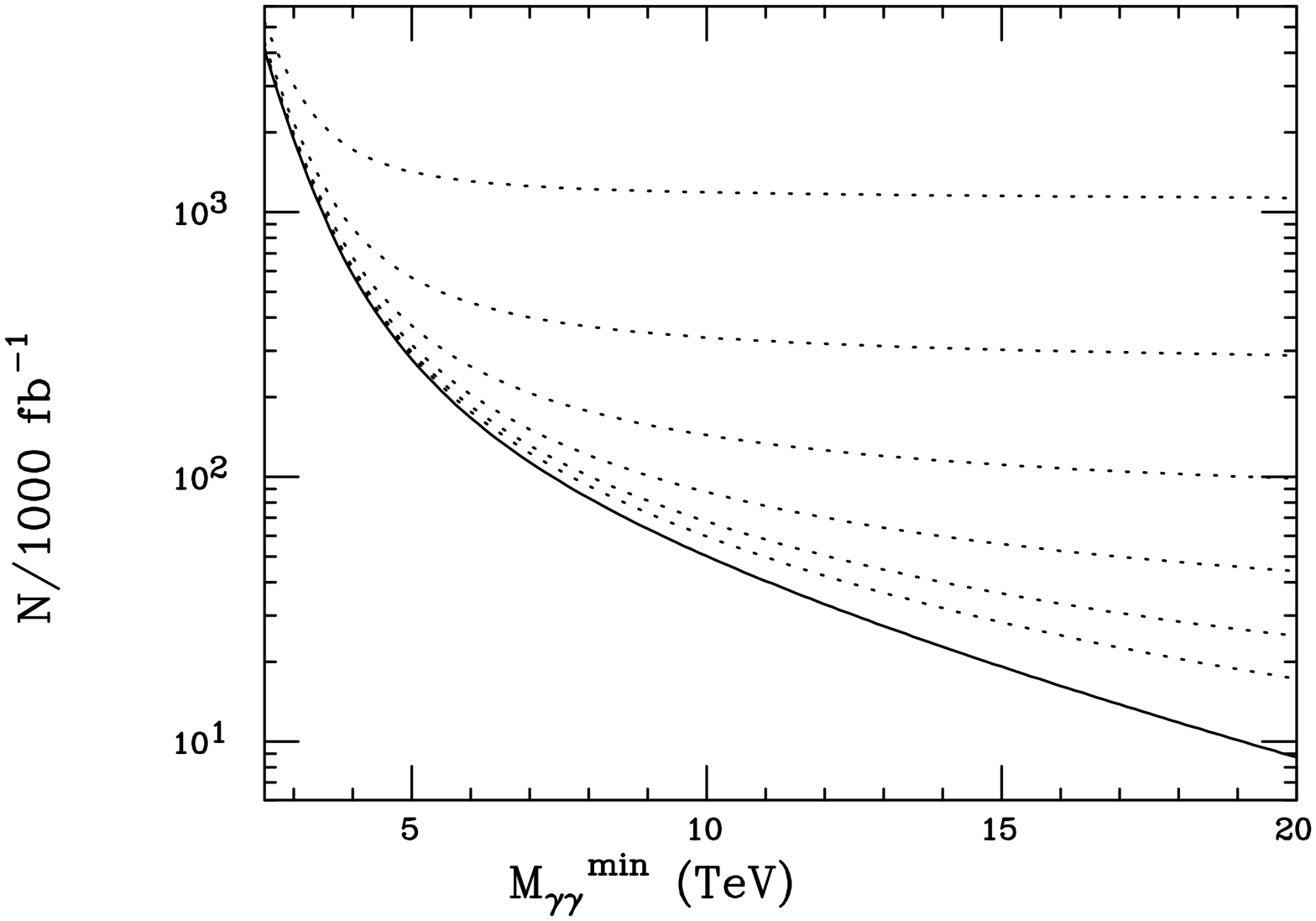,height=9.1cm,width=9.1cm,angle=-90}
\hspace*{-5mm}
\psfig{figure=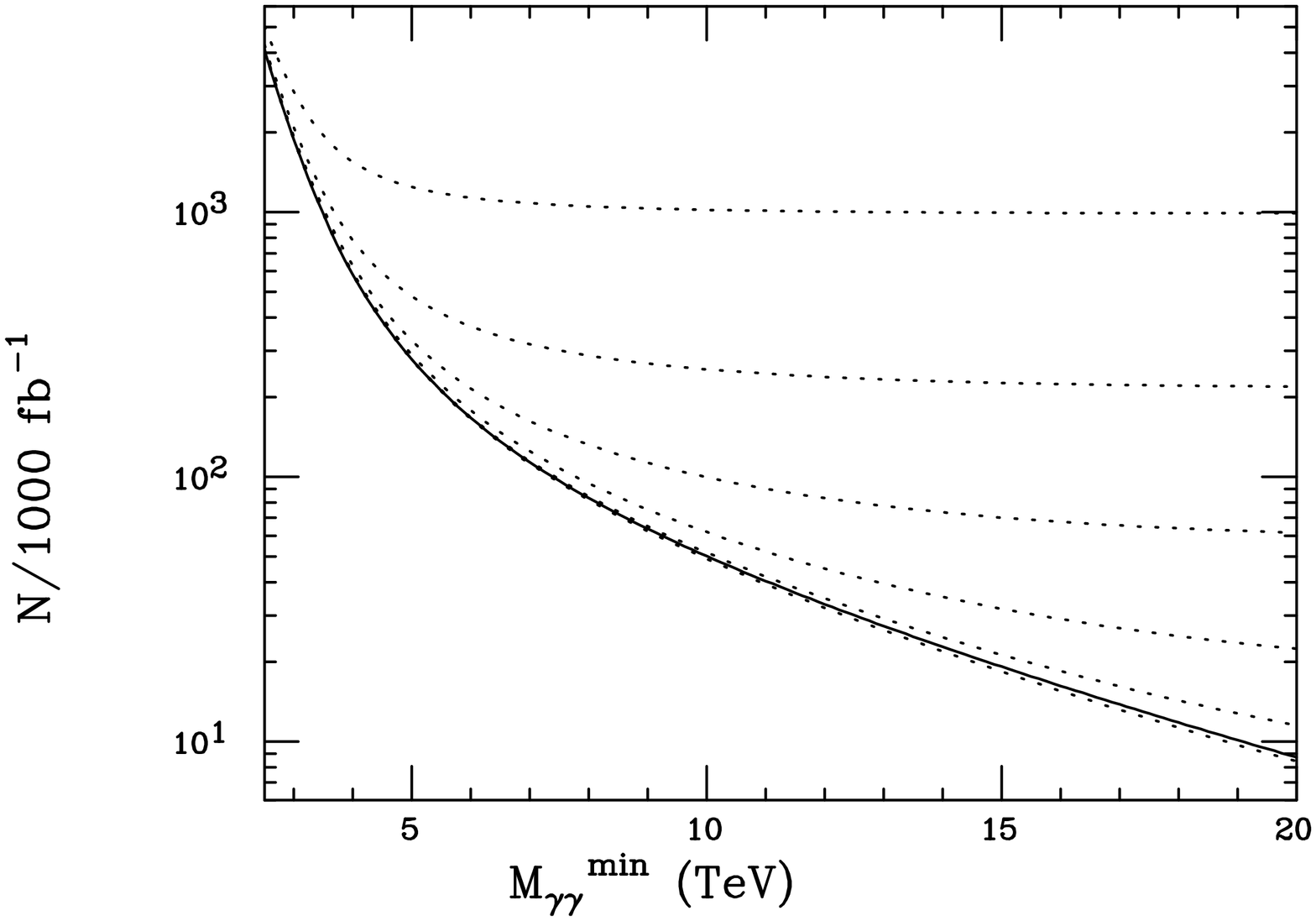,height=9.1cm,width=9.1cm,angle=-90}}
\vspace*{-0.6cm}
\caption{Same as the previous figure but now for the $p\bar p$ collision mode. 
The solid curve is the SM case while the top dotted curve corresponds to 
$\Lambda_+(\Lambda_-)=15$ TeV in the left(right) figure. Each subsequent dotted 
curve corresponds to an increase in $\Lambda_{\pm}$ by 3 TeV.}
\label{200tev2}
\end{figure}
\vspace*{0.1mm}

Assuming that no event excesses are observed, one can ask, \eg,  
for the limits that can be placed on $\Lambda_{\pm}$ as the Tevatron 
integrated luminosity is increased. This was done in our earlier 
analysis{\cite {tgr}} by performing a Monte Carlo study. First  
the $M_{\gamma\gamma}^{min}$ range above 100 GeV was divided into 
nine bins of 50 GeV. Almost all of the sensitivity to finite $\Lambda$ 
lies in this range since for smaller values of $M_{\gamma\gamma}^{min}$ the 
cross section looks very similar to the SM, while for larger values of 
$M_{\gamma\gamma}^{min}$ the event rate is too small to be useful even for 
integrated luminosities well in excess of a few $fb^{-1}$. Events were 
generated using the SM as input and the resulting 
$M_{\gamma\gamma}^{min}$ distribution was fit to a $\Lambda_{\pm}$-dependent 
fitting function. From this, bounds on $\Lambda$ are directly obtainable 
via a $\chi^2$ analysis. In this approach, it was assumed that the 
normalization of the 
cross section for small values of $M_{\gamma\gamma}^{min}$ using experimental 
data will remove essentially all 
of the systematic errors associated with the cross section normalization.  
Hence, only statistical errors are used into the 
fitting procedure. Extending this approach to the case of $\sqrt s$=2 TeV 
Tevatron we obtains bounds on $\Lambda_{\pm}$ in excess of 0.7 TeV as shown 
in Table I. Note that the constraints we obtain on $\Lambda_-$ will generally 
be weaker than those for $\Lambda_+$. Increasing the integrated luminosity at 
the Tevatron by another factor of ten as is proposed in the TeV33 study, will 
most likely push these constraints upwards by approximately 200 GeV.

At the LHC, we find the results presented in Fig. 1b which clearly 
shows that values of $\Lambda$ greater than 2 TeV will be easily probed. 
If one follows the same Monte Carlo approach as above, one obtains very strong 
limits on $\Lambda_{\pm}$. Here the $M_{\gamma\gamma}^{min}$ range above 
250 GeV is divided into ten bins and data is generated as above 
and subsequently fitted to a $\Lambda_{\pm}$-dependent distribution. 
From this analysis we obtain the $95\%$ CL 
bounds of $\Lambda_+>2.83$ TeV and $\Lambda_->2.88$ TeV as displayed 
in Table I. If we increase the integrated 
luminosity to $200 fb^{-1}$, these limits are found to increase 
to $\Lambda_+>3.09$ TeV 
and $\Lambda_->3.14$ TeV. It is interesting to note that if we relax the 
$\eta^{\gamma}$ cuts at the LHC (from 1 to 2.5) we get an increase in 
statistics but a loss of sensitivity. These two effects essentially cancel in 
this case yielding essentially the same bounds as shown in the Table. Note 
that in the case of the LHC we obtain comparable sensitivities to both 
$\Lambda_+$ and $\Lambda_-$.

\begin{table*}[htbp]
\leavevmode
\begin{center}
\label{bounds}
\begin{tabular}{lccccc}
\hline
\hline

Machine& $p_t^{min}$(GeV)& $|\eta_{\gamma,max}|$& $\cal L$& $\Lambda_+$& 
$\Lambda_-$ \\
\hline
TeV                 &  15 & 1 & 2 & 0.75 & 0.71 \\
LHC                 & 200 & 1,2.5 & 100 & 2.8 & 2.9 \\ 
60 TeV ($pp$)       & 500 & 1 & 100 & $\simeq 9.5$ & $\simeq 6.5$ \\
60 TeV ($p\bar p$)  & 500 & 1 & 100 & $\simeq 13.5 $ & $\simeq $ 10.5\\
200 TeV ($pp$)      &1000 & 1 & 1000 & $\simeq 23$ & $\simeq 16$ \\
200 TeV ($p\bar p$) &1000 & 1 & 1000 & $\simeq 33$ & $\simeq 26 $ \\ 
\hline
\hline
\end{tabular}
\caption{$95\%$ CL bounds on the scale of the $q\bar q \gamma\gamma$ 
contact interaction 
at future hadron colliders. Note the greater sensitivity at $p\bar p$ in 
comparison to $pp$ colliders of the same center of mass energy. Here, $\cal L$ 
is the machine integrated luminosity in $fb^{-1}$ and $\Lambda_{\pm}$ are in 
units of TeV.}
\end{center}
\end{table*}

Turning our attention to the higher energy $\sqrt s$=60 and 200 TeV machines, 
we can now explore the differences in sensitivity to $pp$ versus $p\bar p$ 
initial states. Clearly we anticipate larger sensitivities in the $p\bar p$ 
case since the $q\bar q$ process is now proportional to valence times 
valence distributions instead of 
valence times sea.  For the 60 TeV collider our results are shown in 
Figures~\ref{60tev1} and ~\ref{60tev2} assuming an integrated luminosity of 
100 $fb^{-1}$. As in the Tevatron case we see that there is somewhat greater 
sensitivity in the case 
of constructive interference($\Lambda_+$) than in the case of destructive 
interference($\Lambda_-$). Also, as anticipated, the larger $q\bar q$ cross 
section at the $p\bar p$ collider results in substantially enhanced 
sensitivity with an approximate $40\%$ increase in reach for $\Lambda_+$ and 
an approximate $60\%$ increase in reach for $\Lambda_-$. Table I summarizes 
these approximate results; full Monte Carlo studies along the lines discussed 
above for the Tevatron and LHC have not yet been completed. 

At $\sqrt s$=200 TeV this pattern is essentially repeated as can be seen from 
Figures~\ref{200tev1} and ~\ref{200tev2} as well as Table I where an 
integrated luminosity of 1000 $fb^{-1}$ has now been assumed. Note that the 
`reach' in $\Lambda$ sensitivity does not scale linearly with the collider 
energy.

Excess diphoton events should be searched for, not only as narrow peaks in 
$M_{\gamma\gamma}$ signalling the existence of Higgs-like objects, but also 
in the broad contributions to the tails of distributions. Such searches may 
yield valuable information on the existence of new physics.

\section{Acknowledgements}

The author would like to thanks J. Hewett and R. Harris for discussions 
related to this work.

%
\def\MPL #1 #2 #3 {Mod.~Phys.~Lett.~{\bf#1},\ #2 (#3)}
\def\NPB #1 #2 #3 {Nucl.~Phys.~{\bf#1},\ #2 (#3)}
\def\PLB #1 #2 #3 {Phys.~Lett.~{\bf#1},\ #2 (#3)}
\def\PR #1 #2 #3 {Phys.~Rep.~{\bf#1},\ #2 (#3)}
\def\PRD #1 #2 #3 {Phys.~Rev.~{\bf#1},\ #2 (#3)}
\def\PRL #1 #2 #3 {Phys.~Rev.~Lett.~{\bf#1},\ #2 (#3)}
\def\RMP #1 #2 #3 {Rev.~Mod.~Phys.~{\bf#1},\ #2 (#3)}
\def\ZP #1 #2 #3 {Z.~Phys.~{\bf#1},\ #2 (#3)}
\def\IJMP #1 #2 #3 {Int.~J.~Mod.~Phys.~{\bf#1},\ #2 (#3)}

\end{document}